\documentclass[12pt]{article}
\usepackage{graphicx,epsfig,amsmath,amssymb,graphicx} \usepackage{mathrsfs,color,hyperref}

\newcommand{\be}{\begin{equation}}
\newcommand{\ee}{\end{equation}}
\newcommand{\bea}{\begin{eqnarray}}
\newcommand{\eea}{\end{eqnarray}}

\def\({\left(} \def\){\right)}

\begin{document}

\title{\vspace{-1.8in}
\vspace{0.3cm} {Origin of the blackhole information paradox}}
\author{\large Ram Brustein \\
 \hspace{-0.5in} \vbox{
 \begin{flushleft}
  $^{\textrm{\normalsize
\ Department of Physics, Ben-Gurion University,
    Beer-Sheva 84105, Israel}}$
 \\ \small \hspace{1.7in}
    ramyb@bgu.ac.il
\end{flushleft}
}}
\date{}
\maketitle

\begin{abstract}
It is argued that the blackhole information paradox originates from treating the blackhole geometry as strictly classical. It is further argued that the theory of quantum fields in a classical curved space with a horizon is an ill posed problem. If the geometry is allowed to fluctuate quantum mechanically, then the horizon effectively disappears. The sharp horizon emerges only in the classical limit when the ratio of the Compton wavelength of the black hole to its Schwarzschild radius vanishes.  The region of strong gravity that develops when matter collapses to form the blackhole remains visible to the whole of spacetime and has to be described by a microscopic theory of strong gravity.  The arguments imply that the information paradox is demoted from a paradox involving fundamental principles of physics to the problem of describing how matter at the highest densities gravitates.
\end{abstract}
\newpage

\section{Introduction}

The information paradox was posed by Hawking \cite{infopar} after his discovery that  quantum blackholes (BH's) radiate. He realized that when the BH  completely evaporates, nothing is left and that this poses a severe problem for quantum mechanics.

A possible definition of the information paradox (see, for example, \cite{Giddings,Mathur}) is that some tenets of physics have to be abandoned to accommodate the physics of BH's. If  unitarity is the abandoned principle, then  information would be ``lost". If it is locality, then one would have "early information retrieval" from BH's. This is sometimes phrased as ``information must propagate outside the light cone". Finally, the second law of thermodynamics, energy conservation or the closely related  vacuum stability could be thrown under the bus. This is associated with "late information retrieval" and the existence of BH remnants.

The information paradox is a problem of principle, not of practical ability to observe the process of BH evaporation with sufficient precision. This process is often compared to the burning of a book. No one doubts that the information in the book can be retrieved (with enough effort) from the fumes and ashes of the fire. On the other hand, everyone  accepts that this is practically impossible. So the burning of books was never elevated to the level of a paradox or interpreted as an indication that some fundamental physical principles have to be abandoned.

The standard tool for studying quantum BH's is the theory of quantum fields in curved space \cite{BandD}. This theory treats matter quantum mechanically while the background  geometry is treated strictly as classical.  It is possible to improve upon this approximate treatment  by taking into account some quantum mechanical effects of the background. This is done by using the the tools of effective field theory \cite{burgess}. However, after the improvement one ends with a modified, but still  strictly classical geometry. As I will argue, the theory of quantum fields in curved space is particularly vulnerable when the background geometry has a horizon -- an  infinite redshift surface that sharply divides different causally disconnected regions.

The equivalence principle asserts that acceleration and gravitation are equivalent. However, constant acceleration over an infinitely long time leads to the formation of a horizon and to Rindler spacetime. This posses a big problem for quantum mechanics. Is it possible that one gets an observer-dependent thermodynamics? This has led to the introduction of the idea of  ``black hole complementarity" \cite{complementarity1,complementarity2}, that the different observers are mutually exclusive and a global observer does not exist. More recently complementarity has been challenged by the so called ``firewall proposal" \cite{firewall}.

In this paper I will argue that to describe the physics of quantum BH's  no new principles need to be introduced, neither abandoning old ones. What is required is a consistent quantum mechanical treatment of the BH. Quoting from \cite{beknc} ``a black hole must be subject to quantum laws." The notion of quantum matter on a background of a classical geometry with an infinitely sharp horizon is an ill posed quantum mechanical problem. The fundamental physical objects are the global quantum state of the BH and matter and the unitary quantum evolution operator. All parts of the global quantum state are accessible to all, so the issue of observer dependence is not a fundamental one. The thermal nature of black holes is only approximate, becoming exact in the classical limit when the ratio of the Compton wavelength of the BH to its Schwarzschild radius vanishes.

Previous attempts to understand and resolve the information paradox include the S-matrix approach \cite{complementarity2,smatrix2,smatrix3,smatrix4,englert} and the fuzzball approach \cite{fuzz1,fuzz2,fuzz3}. Both share some ingredients and conclusions with the proposed resolution of the information paradox. However, the reasoning and details are quite different.

\section{Supremacy of quantum mechanics over geometry}

Let us recall the well-known example of a test particle in a field of a gravitating distribution of mass in some quantum state \cite{page,Carlip:2008zf} (see also \cite{penrose}.) For concreteness, one can think of a spherical mass distribution. Classically, the particle moves along the geodesics of the Schwarzschild geometry.  How is the motion of the particle described if the spherical mass is in a superposition of two different locations? In this case it is not possible to find  a single classical geometry that the test particle moves on its geodesics. According to the rules of quantum mechanics we are instructed to define a wavefunction for the geometry and the test particle $|\Psi_{M,g}\rangle$. Here the subscript $M$ denotes the matter -- the particle and $g$ stands for the geometry that is created by the distribution of mass.  We are then instructed to calculate expectation values of matter operators $\langle\Psi_{M,g}\left|{\cal O}_M\right|\Psi_{M,g}\rangle$, for example, ${\cal O}_M$ can be the momentum of the particle or its position. On general grounds, one expects that true quantum gravity observables are only boundary observables. However, in the situation that we have described, we do expect approximate local observables as well.

We can treat the test particle as classical. This ia an antipodal approximation to the approximation that is being made in the theory of quantum fields in curved space. The background geometry is treated here as quantum and the matter is treated as classical.
This is, in a way, the inverse of  the ``backreaction" of quantum fields on the background geometry. The term backreaction refers to quantum effects of matter modifying the classical geometry. Here the quantum effects of the geometry modify the dynamics of the matter.

What happens when the distribution of mass has a localized wavefunction around a single position? If the particle is far enough from the mass distribution then it is possible to find an approximate single classical Schwarzschild geometry $g_c$ such that the test particle will follow its geodesics. Then, the wavefunction is only a matter wavefunction which is a function of the classical geometry $\Psi_{M}(g_c)$.
In this case  $\langle\Psi_{M,g}|O_M|\Psi_{M,g}\rangle \sim \langle\Psi_{M}(g_c)|O_M|\Psi_{M}(g_c)\rangle $.

The lesson that we have to learn is that we should follow the rules of quantum mechanics and use them to guide us in making approximations. We should use geometry as the guiding principle only in cases that it does not contradict the rules of quantum mechanics.

\section{Horizon fluctuations effectively destroy the horizon}

In this section I would like to discuss two examples of fluctuating horizons. The conclusions from the discussion will lead us to the next section where it will be argued  that a classical geometry with a horizon (and singularity) is not always a good approximation.  When one computes quantum expectation values of matter operators in the corresponding semiclassical state, the classical geometry should only be considered as a limiting case and the results of using the classical geometry should be interpreted with care.

The first example is inspired by the discussion in \cite{BK} where the fate of a fluctuating infinite redshift surface was considered. The second example is taken from \cite{BH1} where the fate of a fluctuating bifurcating killing horizon was considered. The purpose of the first example is to show that if the horizon is allowed to fluctuate, then it effectively disappears. The second example is more specific and shows that a bifurcating Killing horizon must fluctuate and demonstrates how it disappears.

Here the term ``horizon" has different meaning in different circumstances. In the first example it is used mainly to mean a surface of infinite redshift that separates two regions of spacetime. The surface separates its two sides in the sense that all classical correlations across this surface vanish. In the second example, the specific properties of a bifurcating Killing horizon are used.

\subsection{Fluctuating horizon}

Consider a surface of infinite redshift at some given time $t$. This surface separates the space into two regions - inside and outside, left and right etc. Each region of spacetime is covered by a separate coordinate patch. As a concrete example we can think of the Minkowski and Rindler geometries, or Schwarzschild and Kruskal geometries. Now, imagine that the position of the surface could fluctuate quantum mechanically. What are the implications?

Let us assume for concreteness that the position of the horizon of a Schwarzschild BH could fluctuate quantum mechanically. This means that the semiclassical BH  could be in a general superposition of geometries $ds^2=-\left(1-R/r\right)dt^2 + \frac{1}{1-R/r}dr^2+ r^2 d\Omega_2^2$. The geometries are ``off-shell" so $R\ne 2G M_{BH}$ and can fluctuate independently of the mass. Each of the geometries has a horizon at a different value of $r=R$ and covers a only the part of spacetime outside its horizon $r>R$. The correspondence principle implies that the expectation value of the horizon radius is the classical value $R_S= 2G M_{BH}$ and that if the BH is large,  the fluctuations of the horizon about the mean are small $\Delta R^2/R_S^2 \ll 1$.

Then, we have to describe the BH in terms of a wavefunction of the horizon position $R$. As just argued, we expect that for a large BH the wavefunction is peaked at its classical value ${R}_S$ with some spread $\sigma$,
\begin{equation}
\Psi(R)={\cal N} \ e^{\hbox{$- \frac{\left(R-{R}_S\right)^2}{4 \sigma^2}$}},
\label{radial}
\end{equation}
where ${\cal N}$ is a normalization factor, ${\cal N}^2= 4\pi\int\limits_0^\infty dR R^2  \ e^{\hbox{$- \frac{\left(R-{R}_S\right)^2}{2 \sigma^2}$}}$.

The wavefunction in Eq.~(\ref{radial}) does not necessarily have to be a solution of the true quantum equation that BH's satisfy, however, it will serve to illustrate the point I am trying to make. As we will see in the second example, this wavefunction is in fact an approximate solution for the S-wave mode of a Schwarzschild BH.

A possible way to probe the existence of a horizon for the fluctuating geometry  is to send a light source (which could be classical) towards the horizon. Then one has to check whether light from the source can reach an observer at $r \to \infty$. In the classical geometry, once the probe goes through $r=R_S$, the emitted light can not reach $r \to \infty$. However, the fluctuating geometry is a superposition of geometries. In some of them the position of the horizon is less than the Schwarzschild radius $R < R_S$, so the light emitted inside $r=R_S$ is still outside the horizon.  It follows that with some finite probability, the emitted light can reach $r\to \infty$.

For example, one could calculate the probability $P(R_S)$ that some signal reaches from $r <{R}_S$ to infinity. If we treat the light source as classical and assume that all emitted signals from any of the points outside the horizon can reach infinity, then $P({R}_S)$ is given by
\begin{equation}
P(R_S)=4\pi \int\limits_0^{{R}_S} dR R^2 \left|\Psi(R)\right|^2.
\label{prs}
\end{equation}
Clearly, $P({R}_S)$ is finite. For example, when ${R}_S\gg \sigma$, as expected for semiclassical BH's, $P({R}_S)= 1/2$.   Classically $P({R}_S)$ is strictly vanishing. The probability $P({R})$ decreases rapidly for $R < {R}_S-\sigma$. For $R/{R}_S\ll 1$, $P({R})\sim e^{\hbox{$- \frac{{R}_S^2}{2 \sigma^2}$}}$ is exponentially small. Clearly, for practical calculations it makes sense to approximate an exponentially small probability by zero. However, one has to keep in mind that this is an approximation.

The conclusion is that it is not possible to choose a single classical geometry with a horizon that can reproduce the result in Eq.~(\ref{prs}).  It is in this sense that the horizon fluctuations have effectively destroyed it. If one insists on describing spacetime in with a Penrose diagram, then after taking into account the quantum fluctuations of the geometry, the resulting diagram  has the topology of Minkowski space. Light emitted anywhere in spacetime can reach $r\to \infty$. Penrose diagrams describe only classical motion in conformal coordinates and the causal structure of spacetime and are incapable of encoding information on the quantum state.

The modified diagram taking into account the quantum fluctuations and with the caveats just mentioned is shown in Fig.~1 together  with the standard Pensroe diagram of a Schwarzschild BH.
\begin{figure}[t]
\vspace{-1.9in}\hspace{-0.2in}\scalebox{.55} {\includegraphics[angle=-90]{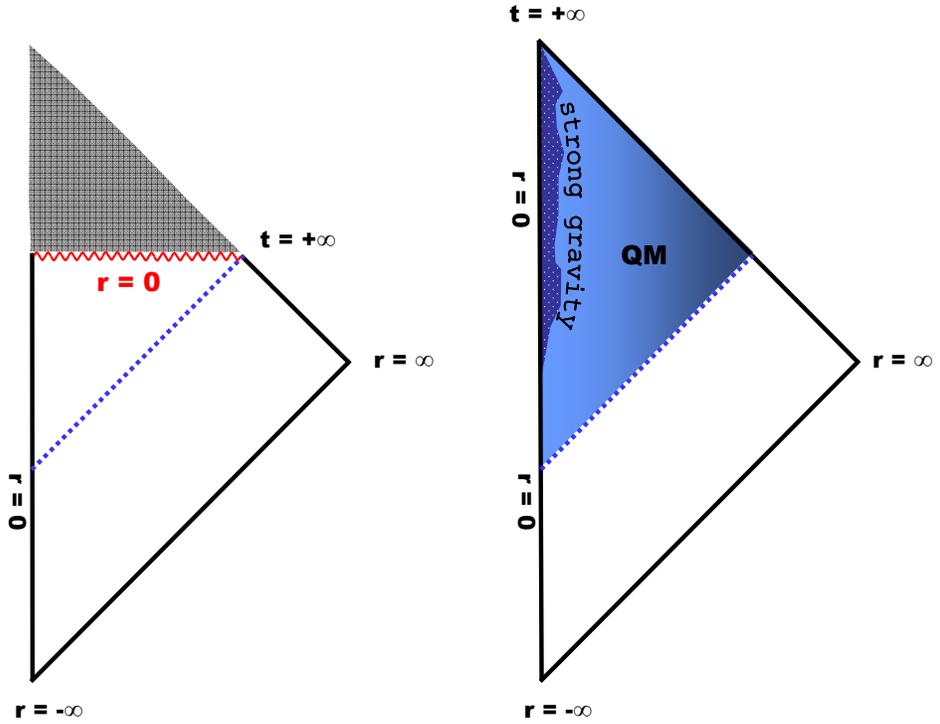}}
\vspace{-.05in}
\caption{Penrose diagrams of blackhole spacetimes. Shown on the left is the standard diagram (evaporation is ignored). The shaded region is excluded. On the right, the diagram after inclusion of horizon quantum fluctuations  is shown. The shaded region near $r=0$ depicts the strong gravity region.}
\end{figure}
The diagram shown in Figure~1 resembles  the Penrose diagram that was obtained in the S-matrix approach by Englert and Spindel \cite{englert}. The reasoning that led to this result, though, is quite different here.

\subsection{Fluctuations of a bifurcating Killing horizon}

Let us now review some of the relevant results from \cite{BH1} and put them in context. In that paper spacetime was foliated by constant $r$ hypersurfaces. The unit normal to the hypersurface is denoted $u_a$ and the hypersurface metric is  $g_{ab} =h_{\alpha\beta}e_{\ a}^\alpha e_{\ b}^\beta-u_a u_b$ where $e_{\ a}^\alpha$ is a basis of tangent vectors to the hypersurface. Greek indices denote the induced coordinates on the
hypersurface. Then a $D-2$ cross section of the hypersurface $\Sigma_{D-2}$ was chosen. The area of  $\Sigma_{D-2}$ is given by $A_{D-2}$.

It was then observed that the gravitational action contains the term
\begin{eqnarray}
\label{action1}
\int d^Dx\sqrt{-g}\frac{\partial\mathscr{L}}{\partial R_{c\alpha\beta d}} \mathcal{L}_u \left(K_{\alpha\beta}u_c u_d\right).
\end{eqnarray}
Equation~(\ref{action1}) implies that the extrinsic curvature $K_{\alpha\beta}u_c u_d$ is canonically conjugate to $\frac{\partial\mathscr{L}}{\partial R_{c\alpha\beta d}}$.  The projections of these variables on the bi-normal vector $\hat{\epsilon}^{\gamma d}$  to  $\Sigma_{D-2}$ satisfy standard Poisson bracket relations:
\begin{eqnarray}
\label{canonical}
\left\{K_{\alpha \gamma}u_b u_d \hat{\epsilon}^{\gamma d}\hat{\epsilon}^{\alpha  b}(x_1), \frac{\partial\mathscr{L}}{\partial R_{ b \alpha \gamma d}}\hat{\epsilon}_{\gamma d}\hat{\epsilon}_{\alpha  b}(x_2)\right\}=(-h)^{-1/2}\delta^{D-1}(x_1\!-\!x_2).
\end{eqnarray}
After some manipulations \cite{BH1} arrives at the Poisson brackets
\begin{eqnarray}
\label{canonical 3}
\left\{\Theta,\frac{1}{2\pi}\,  S_W \right\}=1,
\end{eqnarray}
where $S_W$ is the Wald Noether charge entropy and $\Theta$ is the opening angle at the horizon \cite{carlip}.

The Wald Noether charge entropy \cite{wald1,jkm} in units  such that the BH temperature is equal to $2\pi$ is given by
\begin{equation}
S_W=-2\pi\oint\limits_{H} \frac{\partial\mathscr{L}}{\partial R_{\alpha b  \gamma d}}\hat{\epsilon}_{\gamma d}\epsilon_{\alpha  b}.
\label{walddef}
\end{equation}
In general, it is equal to a quarter of the area of the horizon $A_H$ in units of the effective gravitational coupling $G_{eff}$ \cite{BGH},
\begin{equation}
S_W= \frac {A_H}{4 G_{eff}}.
\label{walddef1}
\end{equation}
For Einstein gravity the effective gravitational coupling is  equal to Newton's constant $G_{eff}=G$ and the Wald entropy is equal to the Bekenstein-Hawking entropy of the BH $S_W=S_{BH}$.

The opening angle at the horizon $\Theta$ is a hyperbolic angle defined in terms of the proper time $\tau=\int\sqrt{-g_{00}}dt$ \cite{carlip},
\begin{equation}
\Theta\equiv\frac{1}{2}\mathcal{L}_u\tau.
\label{deftheta}
\end{equation}
By analytic continuation to Euclidean space one finds for a classical Schwarzschild geometry that
\begin{equation}
\Theta=2 \pi.
\label{theta2pi}
\end{equation}

Both $S_W$ and $\Theta$ should be considered as functionals of the bifurcating surface. As such they are not the standard local observables that we encounter in quantum field theory. They are exactly the kind of observables that are expected in a theory of quantum gravity: global covariant observables that are functionals of the boundaries of spacetime.

Proceeding to quantize the BH, following \cite{carlip}, $\Theta$ and $S_W$ were promoted in \cite{BH1} to quantum operators $\widehat{\Theta}$ and $\widehat{S}_W$.  The Poisson brackets in Eq.~(\ref{canonical 3}) are promoted to commutation relations,
\begin{equation}
\left[\widehat{\Theta},\widehat{S}_W\right]=i \hbar.
\label{commutation}
\end{equation}
The wavefunction of the BH realizes the commutation relations in terms of a differential equation,
\begin{eqnarray}
\label{schrodinger}
\frac{2\pi \hbar}{i}\frac{\partial\Psi}{\partial S_W}=\Theta\Psi(S_W).
\end{eqnarray}

Wavefunctions that are eigenfunctions of $\widehat{\Theta}$ have  a fixed value of $\Theta$. However, any wavefunction with  eigenvalue $\Theta$ that is different from $2\pi$ does not correspond to a geometry that solves the classical Einstein equations. It is an off-shell state. In general, we do not expect states in quantum gravity to correspond to a geometrical spacetime. But, since we are using the semiclassical approximation, the off-shell states do correspond to geometries. In this case to a conical defect geometry without a horizon.

Wavefunctions that are eigenfunctions of $\widehat{S}_W$  are harder to interpret. If $S_W$ is different from the classical value in Eq.~(\ref{walddef1}) then they must, again, correspond to off-shell states of the BH. For these states, the horizon area takes values that differ from the classical values. One can view the deviations from the classical value as resulting from quantum fluctuations of the bifurcation surface (See Section 5 of \cite{BGH}). From this point of view, the fluctuationss originate from angular and radial fluctuations of this  surface. One particular mode of these fluctuations is the S-wave mode when one considers only spherically symmetric deformation of the bifurcation surface. In this case $ S_W= \pi R^2/l_p^2$ and the only effective degree of freedom becomes the horizon position $R$, as in the example in Sect.~(3.1).

Brustein and Hadad found an approximate solution of the wavefunction for a Schwarzschild BH in 4D by using the correspondence principle. The idea is that the expectation values of $\widehat{\Theta}$ and $\widehat{S}_W$ should be equal to their classical values,
\begin{equation}
\label{vevtheta}
\langle\widehat{\Theta}\rangle= 2 \pi
\end{equation}
and
\begin{equation}
\label{vevsw}
\langle\widehat{S}_W\rangle= \frac {A_H}{4 G_{eff}}.
\end{equation}
It follows that the wavefunction  of the BH then cannot be an eigenfunction of either $\widehat{\Theta}$ or $\widehat{S}_W$ because both have well defined expectation values that are dictated by the classical geometry. The uncertainty relation between the two canonically conjugate variables forces the wavefunction to be in a superposition.

Applying the correspondence principle also to the fluctuations of $\widehat{S}_W$ and $\widehat{\Theta}$,  the approximate solution for $\Psi(\Theta)$ was found to be
\begin{eqnarray}
\label{solutiontheta}
\Psi(\Theta)= {\cal N}\ e^{\hbox{$-2\pi\frac{ {R}_S^2 }{l_p^2}\left(\Theta-2\pi \right)^2$}} e^{\hbox{\small $ \frac{i}{\hbar}\pi \frac{ {R}_S^2 }{l_p^2} \Theta$}},
\end{eqnarray}
where ${\cal N}$ is a normalization factor and $l_p$ is the Planck length.

As anticipated, the wavefunction in Eq.~(\ref{solutiontheta}) is a superposition of wavefunctions with fixed $\Theta$.   So $\Theta$ fluctuates and does not have exactly the classical value $2\pi$.  However, as discussed previously, a geometry with any value of $\Theta$ that is different from $2\pi$ does not correspond to a classical geometry with a horizon. The geometry in this case is a horizonless geometry with a conical singularity. We see here explicitly that quantum fluctuations of the geometry, in this case of the hyperbolic angle $\Theta$, have effectively destroyed the horizon. This means that spacetime can not be partitioned into the ``outside" and ``inside" of the horizon. It also means that the inside of the horizon cannot be ignored by outside observers. The possibility of ignoring the inside is the essence of the ability to continue to Euclidean space.  Fluctuations in $\Theta$ effectively destroy the horizon and introduce another boundary at $r=R_S$.

We can see that the horizon gets effectively destroyed by quantum fluctuations from a different perspective by looking in more detail at the wavefunction in the entropy representation.  Here I will write the wavefunction for the S-wave mode of a Schwarzschild BH in 4D Einstein gravity, as discussed above. In this case $ S_W= \pi R^2/l_p^2$. Assuming that $R^2\gg l_p^2$, I obtain
\begin{eqnarray}
\label{solutionfin}
\Psi(R)\sim e^{\hbox{$-\frac{\pi}{2} \frac{\bigl(R-R_S\bigr)^2}{l_P^2}$}}.
\end{eqnarray}
This wavefunction can be compared to  the wavefunction in Eq.~(\ref{radial}). The conclusions from the discussion about the fate of the horizon in Sect.~(3.1) apply also here. I wish to emphasize that the general state of the BH cannot be described by a radial wavefunction and must include also angular fluctuations of the horizon. For such states, it is harder to see explicitly the effective disappearance of the horizon.

For macroscopic BH's the wavefunction is extremely sharp. So for calculating many expectation values it makes sense to approximate it in practice as a $\delta$-function. However, one has to keep in mind that this is an approximation and use it with care when it comes to issues of principle. Another observation is that for $R\ll{R}_S$, $|\Psi(R)|^2\sim e^{-S_{BH}}$. This is the minimal expected magnitude of typical expectation values in a unitary system with entropy $S_{BH}$ \cite{maldacena}.

\section{Origin of information paradox}

As we have seen in the previous sections, the physical quantity that has to be calculated is the expectation value of some matter operator ${\cal O}_M$,
\begin{equation}
\label{vevq1}
\langle{\cal O}_M\rangle= \langle\Psi_{M,BH}\left|{\cal O}_M\right|\Psi_{M,BH}\rangle.
\end{equation}
Here matter includes gravitons in addition to particles from the non-gravitational sectors.

However, in all the standard calculations of the theory of quantum fields in curved space one calculates instead
\begin{equation}
\label{vevq2}
\langle{\cal O}_M\rangle_C = \langle\Psi_{M}(g_c)\left|{\cal O}_M\right|\Psi_{M}(g_c)\rangle,
\end{equation}
where $g_c$ is the classical BH solution of Einstein equations (or its generalization), for example, the Schwarzschild geometry. The subscript $C$ indicates that the geometry is taken as a classical geometry. These calculations are then used to draw conclusions about the properties of quantum black holes.

For a semiclassical Schwarzschild BH, the classical limit is defined in terms of the Compton wavelength of the BH $\lambda_{BH}= \hbar/M_{BH}$ and its Schwarzschild radius $R_S= 2 G M_{BH}$. Using the relation between Newton's constant and the 4D Planck length $l_p=\sqrt{\hbar G}$, one finds that the ratio of the Compton wavelength to the Schwarzschild ratio is given by,
\begin{equation}
\frac{\lambda_{BH}}{R_S}=2 \frac{l_p^2}{R_S^2}.
\label{classlim}
\end{equation}
Here we have used units in which the speed of light is unity. Equation~(\ref{classlim}) generalizes in an obvious way to BH's in higher dimensions. For a macroscopic BH the value of $\lambda_{BH}/R_S$ is extremely small. This ratio appears as a parameter that determines the classicality of the BH in recent discussion of Dvali and Gomez \cite{dvaligomez}, where it is called $1/N$.

The classical limit is defined as the limit when $\lambda_{BH}/R_S\to 0$.   In this limit,
\begin{equation}
\label{vevq3}
\langle{\cal O}_M\rangle_C=
\lim\limits_{\lambda_{BH}/{R_S}\to 0} \langle{\cal O}_M\rangle.
\end{equation}
One can also think about the classical limit as the limit ``$M_{BH}\to\infty$", ``$G\to 0$" and $G M_{BH}$ fixed.
This is the limit to which the theory of quantum fields in curved space applies. In the classical limit for the geometry, the matter can still be treated quantum mechanically. When matter is treated quantum mechanically one finds that the BH radiates and never evaporates. The Hawking radiation is perfectly thermal and so the matter is in a thermal state which is, of course, mixed.

The origin of the information paradox  is in the implicit assumption about what happens away from the limit $\lambda_{BH}/{R_S}\to 0$, when $\lambda_{BH}$ is finite,  so the mass $M_{BH}$ and $G$ are both finite, still keeping the product $M_{BH} G$ fixed. The implicit assumption is that the geometry of the infinite mass case is in some sense, a good approximate geometry also in the finite mass case. We can express this assumption as follows,
\begin{equation}
\label{vevq4}
\langle\Psi_{M,BH}\left|{\cal O}_M\right|\Psi_{M,BH}\rangle_{\biggl|\lambda_{BH}/{R_S}\ {\rm finite}}
\stackrel{{\rm assumption}}{=} \langle\widetilde{\Psi}_{M}(\widetilde{g}_c)\left|{\cal O}_M\right|\widetilde{\Psi}_{M}(\widetilde{g}_c)\rangle + {\cal O}\left(e^{-\lambda_{BH}/{R_S}}\right).
\end{equation}
That there is some (possibly modified) classical geometry $\widetilde{g}_c$ and some (possibly modified) matter wavefunction  $\widetilde{\Psi}_{M}(\widetilde{g}_c)$ that are ``similar" to the classical geometry and matter wavefunction in the infinite mass case in the sense that the two expectation values on both sides of Eq.~(\ref{vevq4}) are equal to all orders in the semiclassical approximation and for all matter operators. The metric $\widetilde{g}_c$ is assumed to posses a horizon that separates spacetime into differen causally disconnected regions.

Neglecting the difference between $\langle{\cal O}\rangle_C$ and  $\langle{\cal O}\rangle$ by taking the classical limit leads to a total suppression of classical correlations of matter across the horizon. In this approximation the Hilbert space becomes a product of the states inside and outside the horizon, the global state of matter becomes a mixed state. Fixing the geometry requires fixing the state of matter, so it becomes impossible to consider other states of matter or discuss quantum coherence and phases for this state and so on.  The physics in this limit is described well in the original paper of Hawking \cite{infopar}.

However, if one wishes to discuss more detailed quantum mechanical question, the quantum nature of the BH must be preserved.
Then the Hilbert space is a single global Hilbert space. The different observers can be related to each other by large redshifts, however, the effective redshift is always finite. And so, by the equivalence principle they have to agree on the results of measurements anywhere in spacetime.

I argue, based on the examples from the previous section, that it is not possible to find a classical geometry $\widetilde{g}_c$ with a horizon such that Eq.~(\ref{vevq4}) is  satisfied  away from the strict classical limit.
Rather, semiclassically
\begin{equation}
\label{vevq77}
\left|\Psi_{M,BH}\right\rangle
= \sum a_i \left|\Psi_{M,i}\right\rangle.
\end{equation}
Here the index $i$ could be   discrete or continuous. If it is continuous, the summation in Eq~.(\ref{vevq77}) turns into an integral.

In the semiclassical approximation, the states $\left|\Psi_{M,i}\right\rangle$ generically correspond to horizonless geometries and so appear to be singular. They do not correspond, as sometimes guessed, to non-singular semiclassical geometries with a horizon. The semiclassical singularity is expected to be resolved by an improved microscopic treatment of the  strong gravitational effects, such as the one provided in some cases by string theory.

Then,
\begin{equation}
\label{vevq8}
\langle\Psi_{M,BH}\left|{\cal O}_M\right|\Psi_{M,BH}\rangle_{\biggl|\lambda_{BH}/{R_S}\ {\rm finite}}=\sum |a_i|^2 \left\langle\Psi_{M,i}\right|{\cal O}_M\left|\Psi_{M,i}\right\rangle.
\end{equation}

To make distinction between the correct matrix element and $\langle{\cal O}_M \rangle_{C}$ more precise we may define a typical  energy scale $E$, appearing in the operator ${\cal O}_M$ and a typical length scale $R$ appearing in the operator ${\cal O}_M$. The operator ${\cal O}_M$ is required to satisfy $E \ll M_{BH}$ to prevent a large back reaction and also $R^2\gg l_p^2$ so that high curvature effects can be neglected. Then,  I argue that corrections for some matter operators can appear already at perturbative orders in the in the semiclassical expansion\footnote{As we have seen, in some cases the corrections are indeed exponentially suppressed rather than power suppressed.}
\begin{equation}
\label{vevq7}
\langle{\cal O}_M\rangle_C= \langle{\cal O}_M\rangle+ O\left(\frac{\lambda_{BH}}{R_S}\right)
\end{equation}
and that this difference does change in a qualitative way the behavior of the expectation values. This will be discussed in more detail in \cite{joeytoappear}.

\section{Demotion of the information paradox to a status of a problem in strong gravity}

The proposed resolution of the information paradox is based on treating the BH as a quantum object. The notion of quantum fields in a background of a classical geometry with an infinitely sharp horizon  is an ill posed quantum mechanical problem. Once arbitrarily small quantum mechanical fluctuation are introduced the horizon effectively disappears. The information paradox is quantum mechanics' way of telling us that.
From this perspective, the answer to the question  ``{\it What is wrong with Hawking's original argument?} " is that the correct quantum mechanical state of a semiclassical  BH can not be described by a classical geometry with a horizon (and a singularity). The answer to the related and often asked question ``{\it How does the information get out of the black hole?} " is simple: it was never in!

The fundamental physical objects  are the global quantum state for the BH and matter and the unitary quantum evolution operator. In the absence of a horizon they are both well-defined. Unitarity has to be accepted as a fundamental property of quantum mechanics. Proving unitarity during black hole formation or evaporation  is extremely hard. This would be the analog of proving the unitary evolution of a burning book.

Quantum fluctuations of the geometry take away the sharp distinction between the inside of the horizon and its outside. Consequently, the region of large curvature formed by the collapsing matter which classically develops into a singularity remains visible to all observers.  The remaining challenge  is then describing this region of strong gravity.  This  issue  may eventually be resolved by quantum mechanics itself, as proposed in \cite{maxbr}. Describing the strong gravity region is similar to the hard and well-known problems that we encounter in dealing with non-linearities of strongly coupled theories.

Thus, the information paradox gets demoted to the contained problem of describing the nature of the strong gravitational fields in a small and very dense region of spacetime.

Other more specific issues remain. For example, when calculating $\langle {\cal O}_M \rangle$ rather than $\langle {\cal O}_M \rangle_C$, are the differences sufficient to ensure the the state of matter inside the horizon is accessible to outside observers with enough precision? Is there enough time prior to the complete evaporation of the BH to perform certain kind of measurements? By how much the emitted radiation differs from a thermal radiation? These are interesting questions that need to be clarified by additional research.

\section*{Acknowledgements}
I would like to thank Merav Hadad, Sunny Itzhaki, Judy Kupferman, Gilad Lifschytz, Joey Medved, Gerard 't Hooft, Maximilian Schmidt-Sommerfeld and Joan Simon for discussions. I thank KITP, University of California, Santa Barbara for hospitality during the program ``Bits, Branes, Black Holes".  The research  was supported by the Israel Science Foundation grant no. 239/10 and by the National Science Foundation under Grant No. NSF PHY11-25915.


\begin{thebibliography}{99}

\bibitem{infopar}
  S.~W.~Hawking,
  ``Breakdown of Predictability in Gravitational Collapse,''  Phys.\ Rev.\ D {\bf 14}, 2460 (1976).  

\bibitem{Giddings}
  S.~B.~Giddings,
  ``The Black hole information paradox,''  hep-th/9508151.  


\bibitem{Mathur}
  S.~D.~Mathur,
  ``Black Holes and Beyond,''  arXiv:1205.0776 [hep-th].  

\bibitem{BandD}
  N.~D.~Birrell and P.~C.~W.~Davies,
  ``Quantum Fields In Curved Space,''  Cambridge, Uk: Univ. Pr. ( 1982) 340p



\bibitem{burgess}
  C.~P.~Burgess,
  ``Quantum gravity in everyday life: General relativity as an effective field theory,''  Living Rev.\ Rel.\  {\bf 7}, 5 (2004)  [gr-qc/0311082].  



\bibitem{complementarity1}
  L.~Susskind, L.~Thorlacius and J.~Uglum,
  ``The Stretched horizon and black hole complementarity,''  Phys.\ Rev.\ D {\bf 48}, 3743 (1993)  [hep-th/9306069].  

\bibitem{complementarity2}
  C.~R.~Stephens, G.~'t Hooft and B.~F.~Whiting,
  ``Black hole evaporation without information loss,''  Class.\ Quant.\ Grav.\  {\bf 11}, 621 (1994)  [gr-qc/9310006].  


\bibitem{firewall}
  A.~Almheiri, D.~Marolf, J.~Polchinski and J.~Sully,
  ``Black Holes: Complementarity or Firewalls?,''  arXiv:1207.3123 [hep-th].  



\bibitem{beknc}
  J.~D.~Bekenstein,
  ``The quantum mass spectrum of the Kerr black hole,''
  Lett.\ Nuovo Cim.\  {\bf 11 } (1974)  467.




\bibitem{smatrix2}
  D.~Amati, M.~Ciafaloni and G.~Veneziano,
  ``Towards an S-matrix description of gravitational collapse,''  JHEP {\bf 0802}, 049 (2008)  [arXiv:0712.1209 [hep-th]].  

\bibitem{smatrix3}
  S.~B.~Giddings and R.~A.~Porto,
  ``The Gravitational S-matrix,''  Phys.\ Rev.\ D {\bf 81}, 025002 (2010)  [arXiv:0908.0004 [hep-th]].  

\bibitem{smatrix4}
  G.~'t Hooft,
  ``Quantum gravity without space-time singularities or horizons,''  arXiv:0909.3426 [gr-qc].  


\bibitem{englert}
  F.~Englert and P.~Spindel,
  ``The Hidden horizon and black hole unitarity,''  JHEP {\bf 1012}, 065 (2010)  [arXiv:1009.6190 [hep-th]].  

\bibitem{fuzz1}
  S.~D.~Mathur,
  ``The Fuzzball proposal for black holes: An Elementary review,''  Fortsch.\ Phys.\  {\bf 53}, 793 (2005)  [hep-th/0502050].  

\bibitem{fuzz2}
  I.~Bena and N.~P.~Warner,
  ``Black holes, black rings and their microstates,''  Lect.\ Notes Phys.\  {\bf 755}, 1 (2008)  [hep-th/0701216].  

\bibitem{fuzz3}
  K.~Skenderis and M.~Taylor,
  ``The fuzzball proposal for black holes,''  Phys.\ Rept.\  {\bf 467}, 117 (2008)  [arXiv:0804.0552 [hep-th]].  



\bibitem{page}
  D.~N.~Page and C.~D.~Geilker,
  ``Indirect Evidence for Quantum Gravity,''  Phys.\ Rev.\ Lett.\  {\bf 47}, 979 (1981).  


\bibitem{Carlip:2008zf}
  S.~Carlip,
  ``Is Quantum Gravity Necessary?,''  Class.\ Quant.\ Grav.\  {\bf 25}, 154010 (2008)  [arXiv:0803.3456 [gr-qc]].  

\bibitem{penrose}
  R.~Penrose,
  ``On gravity's role in quantum state reduction,''  Gen.\ Rel.\ Grav.\  {\bf 28}, 581 (1996).  


\bibitem{BK}
  R.~Brustein and J.~Kupferman,
  ``Black hole entropy divergence and the uncertainty principle,''  Phys.\ Rev.\ D {\bf 83}, 124014 (2011)  [arXiv:1010.4157 [hep-th]].  

\bibitem{BH1}
  R.~Brustein and M.~Hadad,
  ``Wave function of the quantum black hole,''  arXiv:1202.5273 [hep-th].  

\bibitem{carlip}
  S.~Carlip and C.~Teitelboim,
  ``The Off-shell black hole,''
  Class.\ Quant.\ Grav.\  {\bf 12}, 1699 (1995)
  [arXiv:gr-qc/9312002].

\bibitem{wald1}
  R.~M.~Wald,
  ``Black hole entropy is the Noether charge,''
  Phys.\ Rev.\  D {\bf 48}, 3427 (1993)
  [arXiv:gr-qc/9307038].

\bibitem{jkm}
  T.~Jacobson, G.~Kang and R.~C.~Myers,
  ``On black hole entropy,''  Phys.\ Rev.\ D {\bf 49}, 6587 (1994)  [gr-qc/9312023].  

\bibitem{BGH}
  R.~Brustein, D.~Gorbonos and M.~Hadad,
  ``Wald's entropy is equal to a quarter of the horizon area in units of the effective gravitational coupling,''  Phys.\ Rev.\ D {\bf 79}, 044025 (2009)  [arXiv:0712.3206 [hep-th]].  

\bibitem{maldacena}
  J.~M.~Maldacena,
  ``Eternal black holes in anti-de Sitter,''  JHEP {\bf 0304}, 021 (2003)  [hep-th/0106112].  



\bibitem{dvaligomez}
  G.~Dvali and C.~Gomez,
  ``Black Hole's Quantum N-Portrait,''  arXiv:1112.3359 [hep-th].  

\bibitem{joeytoappear}
R.~Brustein and A.~J.~M.~Medved, to appear.

\bibitem{maxbr}
R.~Brustein and M.~Schmidt-Sommerfeld, ``Universe explosions", to appear.


\end{thebibliography}
\end{document}